\documentclass[epj,referee]{svjour}
\usepackage{graphics}
\begin{document}
\title{Dynamical and stationary critical behavior of the Ising ferromagnet in a thermal gradient
}
%\subtitle{Do you have a subtitle?\\ If so, write it here}
\author{Juan Muglia\inst{1} \and Ezequiel V. Albano\inst{1,2}% etc
% \thanks is optional - remove next line if not needed
%\thanks{\emph{Present address:} Insert the address here if needed}%
}                     % Do not remove
%
%\offprints{}          % Insert a name or remove this line
%
\institute{Instituto de F\'{\i}sica de L\'{i}quidos y Sistemas Biol\'{o}gicos.  (IFLYSIB),
CCT La Plata. CONICET, UNLP.
Calle 59 No. 789, (1900) La Plata, ARGENTINA. FAX: 0054-221-4257317 
\and 
Departamento de F\'{\i}sica, Facultad de Ciencias Exactas. UNLP; La Plata Argentina.
E-mail: ezequielalb@yahoo.com.ar
}

\authorrunning{Muglia and Albano}
\titlerunning{The Ising magnet in a thermal gradiente}
\date{Received: date / Revised version: date}
% The correct dates will be entered by Springer
%
\abstract{
In this paper we present and discuss results of Monte Carlo numerical simulations of the two-dimensional 
Ising ferromagnet in contact with a heat bath that intrinsically has a thermal gradient. The extremes of 
the magnet are at temperatures $T_1<T_c<T_2$, where $T_c$ is the Onsager critical temperature. 
In this way one can observe a phase transition between an ordered phase ($T<T_c$) and a disordered 
one ($T>T_c$) by means of a single simulation. \\ 
By starting the simulations with fully disordered initial configurations with magnetization $m\equiv 0$ 
corresponding to $T=\infty$, which are then suddenly annealed to a preset thermal gradient, we 
study the short-time critical dynamic behavior of the system. Also, by setting a small initial 
magnetization $m=m_0$, we study the critical initial increase of the order parameter. Furthermore,
by starting the simulations from fully ordered configurations, which correspond to the ground state 
at $T=0$ and are subsequently quenched to a preset gradient, we study the critical relaxation dynamics 
of the system. Additionally, we perform stationary measurements ($t\rightarrow\infty$) that
are discussed in terms of the standard finite-size scaling theory.\\
We conclude that our numerical simulation results of the Ising magnet in a thermal gradient, which are rationalized
in terms of both dynamic and standard scaling arguments, are fully consistent with well established results obtained
under equilibrium conditions.\\     
%\vskip 0.5 true cm
%\vskip 0.5 true cm
%\vskip 0.5 true cm
} %end of abstract
\maketitle
\section{Introduction}
\label{intro}
Let us assume that a physical system undergoes a smooth transition between two distinctly characteristic 
phases when a certain control parameter (e.g., $T=$ temperature, $P=$ pressure, $\mu=$ chemical potential, 
etc.) is finely tuned around a critical value. This physical situation is often addressed by means of
analitical and numerical methods where the control parameter asuumes a fixed value.
However, here we considered an alternative approach by assuming that along a given spatial direction, 
the system undergoes a well established gradient of the control parameter, such that the critical point 
lies between the extreme values, e.g., $T_1< T_c<T_2$, where one has a thermal gradient between temperatures 
$T_1$ and $T_2$, and the critical temperature $T_c$ is within 
that range \cite{dgrad0,dgrad1,dgrad2,dgrad3,jsm}. 
Of course, a standard reversible system, usually studied under equilibrium conditions, will be out of 
equilibrium under the gradient constraint applied to the control parameter. However, we will show below that 
this situation is no longer a shortcoming for the application of methods and theories already developed 
for the study of equilibrium systems.
In order to fix ideas, in this paper we will study the dynamical and stationary behavior of the Ising 
model for a two-dimensional magnet\cite{isingorig} in a thermal gradient. The
proposed study not only possess interesting theoretical challenges, but it may also be useful in connection to recent
experimental work aimed to characterize films in general, and magnetic films in particular, which are obtained under
thermal gradient conditions imposed on the substrate.
In fact, since the temperature is a key parameter for the relevant properties of thin films, several experiments have
focused on the influence of a temperature gradient on film growth, e.g., Tanaka et al. reported studies of magnetic
Tb-Te films obtained under thermal gradient conditions\cite{tanaka}. Also, Schwickert et al.\cite{aleman} have introduced
the temperature wedge method where a temperature gradient of several hundred Kelvin was established across a substrate
 during the co-deposition of Te and Pt. Other experiments performed by Xiong et al.\cite{xiong} involve nanostructures
obtained by using a gradient temperature.

So, within the broad context discussed above, we now focus our attention on the aim of this paper, 
which is to perform an extensive numerical study of the two-dimensional Ising
ferromagnet in a thermal gradient. In previous numerical studies performed by various 
authors\cite{dgrad0,dgrad1,dgrad2,dgrad3,jsm}, the two extremes of the magnet are considered in contact 
with different thermal baths, e.g. the left-hand and the right-hand
side extremes are at fixed temperatures $T_1$ and $T_2$, respectively. 
In order to study this particular out-of-equilibrium situation,
varoius algorithms can be used, e.g. 
the Creutz algorithm \cite{dgrad0,creutz}, the Kadanoff-Swift move \cite{KS}
the Q2R rule \cite{Q2R}, the KQ rule \cite{dgrad2}, a recently introduced 
microcanonical dynamics \cite{jsm}, etc.
In this way, the system naturally evolves into a stationary
state and a thermal gradient is established between its extremes.
Also, the surfaces in the direction perpendicular to the one where
the gradient is observed are surrounded by an adiabatic
wall. In particular, Neek-Amal et al. \cite{dgrad1} have reported
the ocurrence of an almost linear temperature gradient (c.f. figure 1 of reference  \cite{dgrad1}),
for the case of the $d= 2$ Ising model, by using the Creutz algorithm.
On the other hand, the mean-field treatment given by the well known
Fourier equation \cite{kit}, namely

\begin{eqnarray}\label{laplace}
\frac{\partial^2 T(x,t)}{\partial x^2}-K\;\frac{\partial T(x,t)}{\partial t}=0,\nonumber\\
T(0,t)=T_1,\;T(L,t)=T_2,\nonumber
\end{eqnarray}
\noindent where $T(x,t)$ is the temperature along the solid and $K$ is the 
(constant) thermal conductivity, predicts a linear temperature gradient between both baths.
In fact, we assumed a solid of length $L$ where the gradient is applied, while the remaining directions 
are irrelevant. Then, in the $t\rightarrow\infty$ limit, the time-dependent
contributions to the solution become negligible, and one gets the linear gradient, namely,
\begin{equation}\label{gradi}
T(x)=T_1+\frac{(T_2-T_1)}{L}\;x.
\end{equation}

It is worth mentioning that the above simulation studies \cite{dgrad0,dgrad1,dgrad2,jsm} of 
the Ising system are precisely focused on the calculation of the thermal conductivity, 
among other relevant observables, such as the energy profiles of the system’s transversal sections.
Also, in all cases the discussed results were obtained under stationary conditions. 
However, simulation results that are in contrast to the mean-field theory, 
clearly show that the thermal conductivity is not a constant in
these systems since it depends on the local
temperature (or energy) \cite{dgrad0,dgrad1,dgrad2,jsm},
hence the temprature may not necessarily grow
linearly along the heat propagation
direction. In fact, in the presence to thermal fluctuations it is expected that
transport properties may exhibit non-trivial spatial patterns due to the 
non-equilibrium nature of the system under study \cite{jsm}, and 
under these conditions the thermal conductivity depends on 
the temperature (apart from other relevant parameters of
the model). In particular, for the Ising system in a thermal gradient,
Agliari et al. \cite{jsm} reports that $K(T)$ exhibits a peak slightly 
above criticality, whose position is independent of both the lattice size
and the temperature difference between the extrems of 
the sample. Therefore, the system that we study in the present paper is
not simply an Ising model where different temperatures are imposed
at the bundaries, but it describes a system where the temperature
gradient is present in the heat
bath itself and the transport properies of the system are determined
also by the bath and not only by the Ising dynamics. 
While the linear gradient used in our calculations is motivated by
the fact that we considered a sample in contact with a thermal bath
that intrinsically has a gradient, that assumption can also be
supported by a physical argument. In fact, to define a thermal
conductivity there must exist mechanisms such that the degrees of freedom
(e.g. the phonons) dominate the thermal properties and spins
locally thermalize to the temperature of the lattice.
In the absence of such mechanisms the phonons at
one end of the crystal will not be in thermal equilibrium
at a temperature $T_2$, and those at the other end in equilibrium
at $T_1$ \cite{kit}. Then, if we assume that the lattice
has a constant conductivity, at least in the temperature range where
the magnetic transition takes place, we expect a linear growth of
the temperature when a temperature difference is apllied to the
extrems. 

In the present paper we not only undertake a completely different approach than those used in previous studies,  
as already anticipated, but we are also interested in both the dynamic and the stationary behavior of a 
broad spectra of physical observables. In fact, 
we consider that the Ising ferromagnet is already in contact
with a thermal bath that intrinsically has a thermal gradient along the $x-$direction, as described 
by equation (\ref{gradi}). In this way each column of coordinates $x_i=i$, with $i=1,2,....,L$ 
of our two-dimensional (lattice) system, is in equilibrium with the thermal
bath at temperature $T(x_i)$. Under this condition we apply the standard Metropolis dynamics by considering 
the proper temperature for each column. So, the whole system reaches a stationary, non equilibrium regime, 
with a net flux of energy from the hotter to
the colder extremes at temperatures $T_2$ and $T_1$, respectively.

As already anticipated, we are interested in the study of the dynamic behavior of the system, 
i.e., by addressing both the relaxation dynamics and the short-time dynamics\cite{albano}, 
as well as the stationary behavior that is established in the long-time regime.
By using the dynamic scaling theory \cite{albano} and the finite-size scaling theory \cite{binder},
both developed for the analysis of critical behavior of samples in homogeneous thermal baths,
we are able to rationalize all measurements performed in our thermal gradient 
system. In this way we determined not only the critical temperature but also relevant critical
exponents such as those of the order parameter ($\beta$), the susceptibility ($\gamma$), the
correlation length ($\nu$), etc. We also show that the hyperscaling relationship given by
$d^{*} -2\beta/\nu = \gamma/\nu$ holds, with an effective dimension $d^{*} = 1$, which reflects the 
fact that the susceptibility is measured as the fluctuations of the order parameter (magnetization)
per unit of length along the direction perpendicular to the gradient. 

The paper is organized such that in Section \ref{dos} we provide the description of the model and 
the simulation method, in order to allow for a smooth introduction of the theoretical background 
in Section \ref{tres}. In section \ref{cuatro} we present and discuss the
results. Finally, our conclusions are stated in Section \ref{cinco}.

\section{BRIEF DESCRIPTION OF THE ISING MAGNET IN A THERMAL GRADIENT AND THE SIMULATION METHOD}
\label{dos}

We performed simulations of the Ising model in the two-dimensional square lattice with a rectangular geometry,
with a Hamiltonian given by
\begin{equation}\label{hamil}
H=-J\sum_{\langle ij,i^{\prime}j^{\prime}\rangle}\,s_{i,j}s_{i^{\prime},j^{\prime}},
\end{equation}
\noindent where the spin variables can assume two values, i.e. $s_{ij}=\pm 1$, $J>0$ is the coupling 
constant that is set positive in order to account for ferromagnetic interactions, and summation 
runs over nearest-neighbor sites as indicated by the symbol
$\langle ij,i^{\prime}j^{\prime}\rangle$. The Ising system is an archetypal model
for the study of second-order equilibrium phase transitions. In fact, it is well
known that the local interactions of the Hamiltonian given by equation (\ref{hamil})
lead to a macroscopically observable transition between an ordered ferromagnetic
phase and a disordered paramagnetic one, which in $d=2$ dimensions takes place at the Onsager critical 
temperature $T_c=\frac{2J}{k_B\ln(1+\sqrt{2})}$\cite{onsager}, where $k_B$ is the
Boltzmann constant (hereafter we report all temperature values in units of $J/k_B$).

In this work we placed the Ising ferromagnet of size $L\times M$ in contact with a thermal bath 
that intrinsically has a thermal gradient along the $L-$direction, as given by equation (\ref{gradi}). 
Therefore, each column of length $M$ located at $x=i$, with $1\leq i\leq L$, is in thermal equilibrium 
with the bath temperature $T(i)=T_2+(T_2-T_1)i/L$, where $T_1$ and $T_2$ are
the temperatures on the left- and right-hand sides of the magnet, respectively. Since we are interested 
in the study of the critical behavior, the temperatures are selected such that $T_1<T_c<T_2$, so that 
a phase transition between the ordered phase on the left-hand side and the disordered one at the 
right-hand side can be observed in a single simulation run. As imposed by the simulation
geometry used we take periodic boundary conditions in the direction perpendicular to the gradient, while
open boundary conditions are adopted for the extremes of the sample in the direction parallel to the gradient.

In order to perform the Monte Carlo simulation, a spin $s_{ij}$ selected at random and the change 
of the energy ($\Delta H$) involved in a flipping attempt $s_{ij}\rightarrow -s_{ij}$ are evaluated 
by using the Hamiltonian given by equation (\ref{hamil}).
Now, due to the applied thermal gradient, the flipping probability given by the Metropolis dynamics
$W(s_{ij}\rightarrow -s_{ij})=exp(-\Delta H/k_B T(i))$ depends on the $i$th column ($1\leq i\leq L$) 
where the selected spin is located. Subsequently, the standard Monte Carlo procedure is implemented.

The upper panel of figure 1 shows a typical snapshot configuration obtained
under stationary conditions by using the above-described procedure, while
the lower panel shows a sketch of the linear gradient applied to the sample.
%%%%%%%%%%%%%%%%%%%%%%%%%%%%
\vskip 1.0 true cm
\begin{figure}\label{fig1}
\centerline{
\resizebox{0.65\columnwidth}{!}{%
  \includegraphics{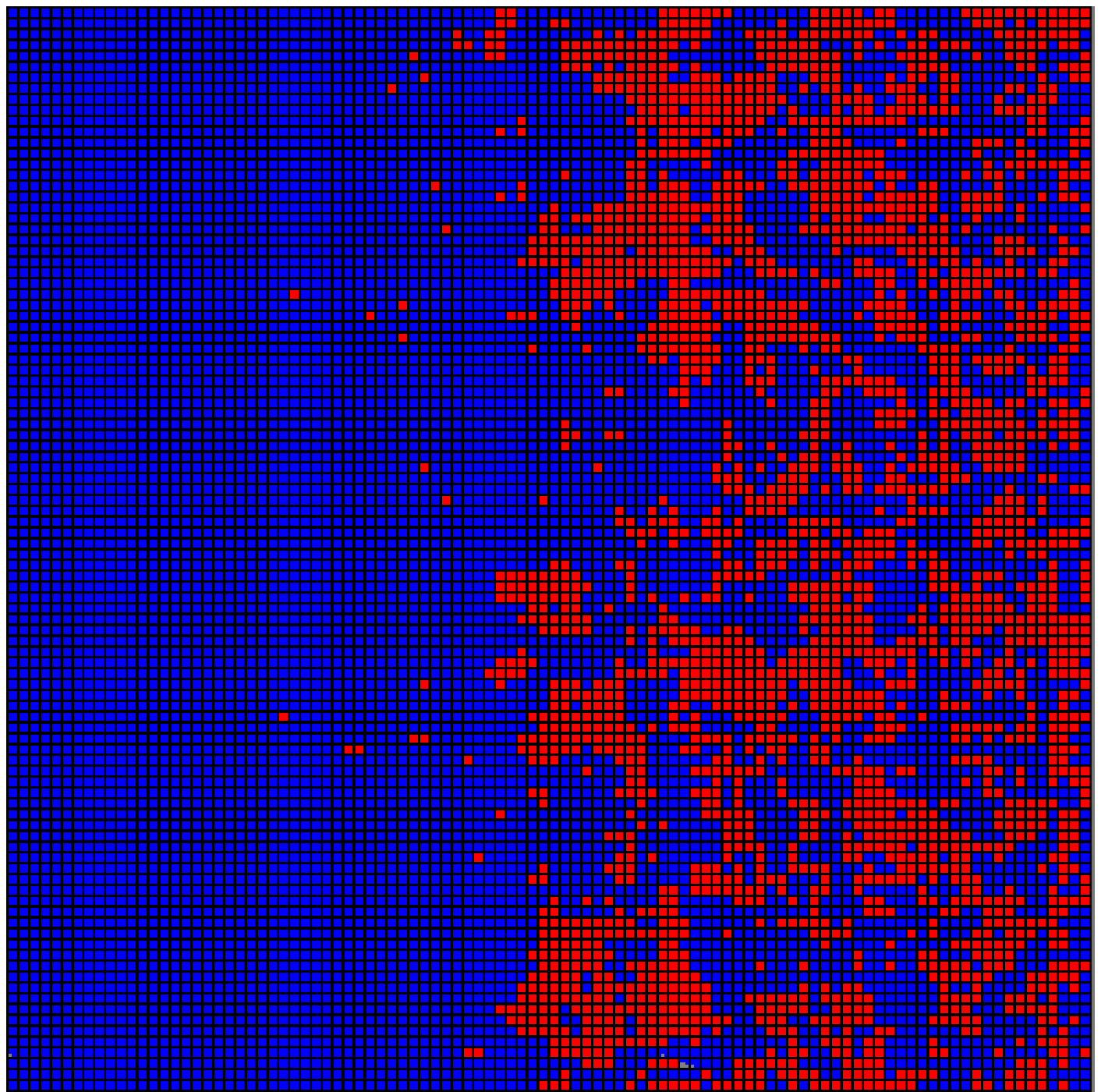}
}}
\centerline{
\resizebox{0.75\columnwidth}{!}{%
  \includegraphics{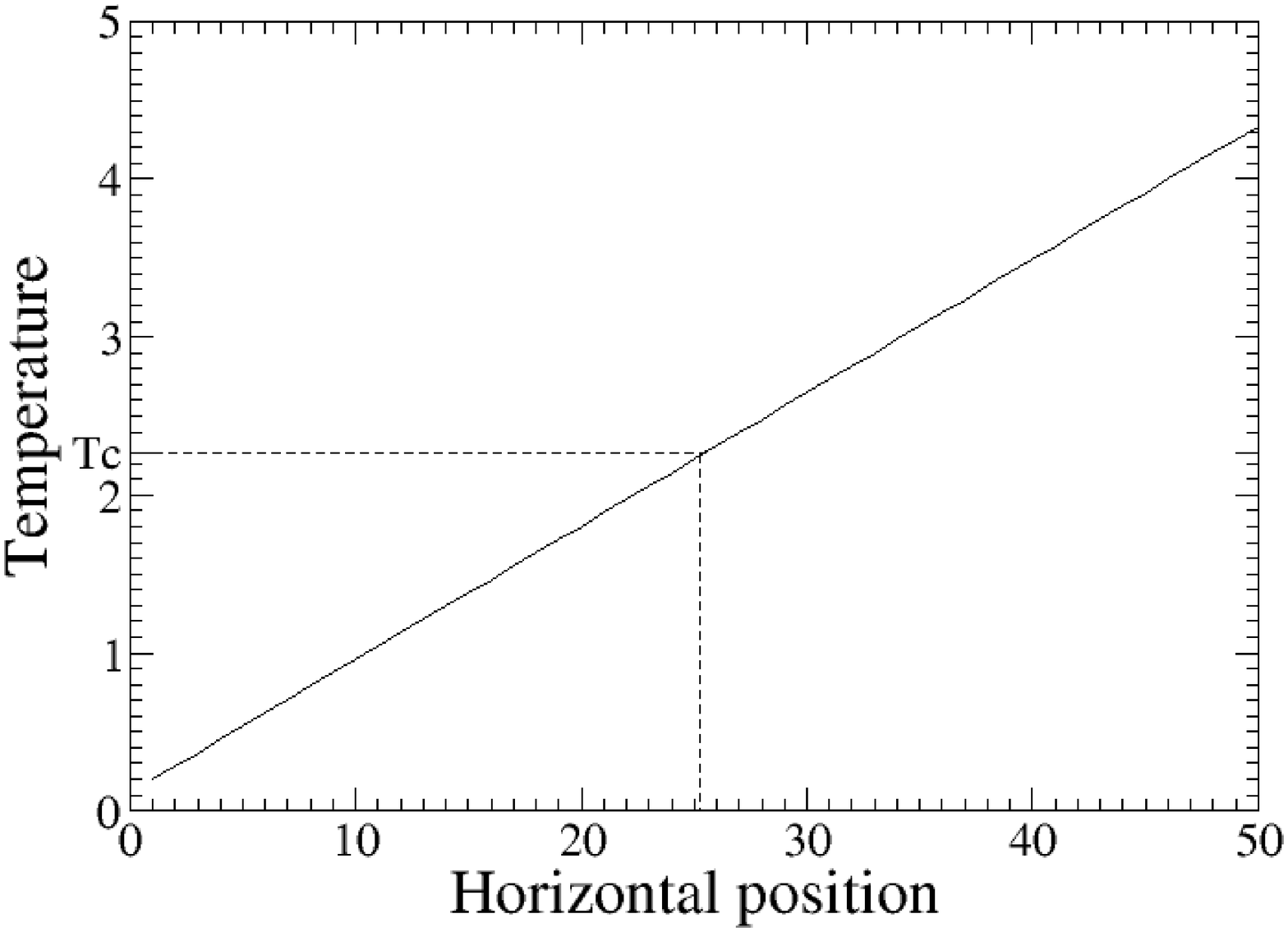}
}}
\caption{(Color online) The upper panel shows a typical snapshot configuration of a two-dimensional 
Ising ferromagnet in contact with a thermal bath that intrinsically has a linear temperature gradient 
that grows from $T_1=0.2$ (left-hand side) to $T_2=4.338$ (right-hand side). Up and down spins are 
shown in blue and red, respectively. The lower panel is a sketch of the applied
gradient, showing the critical region around $T_c\simeq 2.269$ by means of a dotted line. 
The critical interphase between the ordered phase
(left-hand side) and the disordered one (right-hand side) can be observed in the snapshot. 
More details in the text.}
\end{figure}
\vskip 2.5 true cm
%%%%%%%%%%%%%%%%%%%%%%%%%%%%
Due to the applied thermal gradient, all relevant physical observables depend on the coordinate running
in the direction of the gradient itself, so one essentially has to measure profiles of the observables, such as
the nth moment of the order parameter (magnetization in this case) is given by
\begin{equation} \label{magn}
m_i^{(n)}\,(T(i),x,t)=\bigg{(}\frac{1}{M}\,\sum_{j=1}^M\;s_{i,j}(t)\bigg{)}^n,
\end{equation}
where the dependence on the gradient axis ($1\leq i \leq L$) is explicitly indicated. Here, we have also included
the time dependence, where a Monte Carlo time unit involves the random selection of $L\times M$ spins,
as usual.

Furthermore, other measured observables are the susceptibility profiles, evaluated as the magnetization fluctuations,
i.e.,
\begin{equation} \label{chi}
\chi_i(T(i),t)=\frac{1}{M}\,[m_i^{(2)}(T(i),t)-(m_i^{(1)}(T(i),t))^2];
\end{equation}
and the second- and the fourth-order cumulants
\begin{equation}
U_2(T(i),t)=\frac{m_i^{(2)}(T(i),t)}{(m_i^{(1)}(T(i),t))^2}-1,
\label{u2}
\end{equation}

\noindent and

\begin{equation}
U_4(T(i),t)=1-\frac{m_i^{(4)}(T(i),t)}{3(m_i^{(2)}(T(i),t))^2},
\label{u4}
\end{equation}
respectively.

In order to understand the behavior of spin-spin
correlations in a system in the presence of a temperature gradient, 
we also measured the spin-spin spatial correlation function in the direction perpendicular to the thermal
gradient, defined for each column $i$ at temperature $T_i$ as
\begin{equation}\label{ge}
g(T_i,r)=\frac{1}{(M-r)}\,\sum_{j=1}^{M-r}\;(\langle s_{i,j}s_{i,j+r}\rangle-\langle s_{i,j}\rangle\langle s_{i,j+r}\rangle),
\end{equation}
where $r$ is the distance between spins. Furthermore, we also measured the correlation function in
the direction parallel to the thermal gradient, by fixing the origin just at $T_c$.

For the sake of completeness it is worth mentioning that for dynamic simulations the time
dependence of the already defined observables is obtained by averaging over a (large) 
number of different realizations ($N_R$) performed with different sets of random
numbers and physically equivalent initial conditions. Also, measurements corresponding to
the stationary regime are performed by taking averages after a large number ($N_c$) of different configurations,
which are obtained after disregarding a (large) number ($N_D $)
of Monte Carlo time steps, in order to allow for the system stabilization.

\section{THEORETICAL BACKGROUND}
\label{tres}

By considering a fully disordered sample but with a small initial magnetization ($m_0$), the general scaling behavior
of the n-th moment of the magnetization that it is expected to hold for temperatures close to the critical region,
such that $\epsilon\equiv\frac{(T(i)-T_c)}{T_c}\simeq 0$, is given by\cite{albano},
\begin{equation}\label{escala}
m^{(k)}(\epsilon,t,L)=b^{-\frac{k\beta}{\nu}}\tilde m^{(k)}(b^{-z}t,b^{\frac{1}{\nu}}\epsilon,b^{-1}L,b^{x_0}m_0),
\end{equation}
where $\beta$ and $\nu$ are the order parameter and the correlation length (static) critical exponents, $z$ is the
dynamic exponent, and $b$ is a suitable scaling parameter\cite{albano}. Accordingly to the short-time dynamic scaling
theory, $x_0$ is a new exponent that governs the early-time scaling behavior of the moments of the initial
magnetization. Now, by setting $b\equiv t^{1/z}$, just at the critical point
 ($\epsilon\equiv 0$\cite{foot0}), and for the
short-time regime such that the correlation length $\xi(t)\sim t^{1/z}$ is smaller than the typical lattice side
($\xi<L,M$) but slightly larger than the lattice spacing, one has that for $k=1$ equation (\ref{escala}) becomes
\begin{equation}\label{theta}
m(t)\sim t^{\theta},
\end{equation}
where the exponent $\theta=\frac{x_0-\beta/\nu}{z}$ is the scaling exponent
of the initial increase of the magnetization and equation (\ref{theta})
holds for $t^{x_0/z}\ll 1$\cite{janssen}.

Furthermore, for $\epsilon=0$ and $m_0\equiv 0$ the scaling behavior of the fluctuations of the 
second moment of the magnetization, which give the
susceptibility according to equation (\ref{chi}), is given by\cite{albano}
\begin{equation}\label{chiescala}
\chi(t)\sim t^{(d-2\beta/\nu)/z}.
\end{equation}
It is worth mentioning that if the hyperscaling relationship $\nu d-2\beta=\gamma$ holds, one has the 
well-known dependence $\chi(t)\sim t^{\gamma/\nu z}$,
where $\gamma$ is the susceptibility exponent.

On the other hand, by starting from a well ordered initial configuration, e.g., for $m_0=1$, 
the scaling approach given by equation (\ref{escala}) for $k=1$ becomes
\begin{equation}\label{magescala}
m(\epsilon,t) \sim t^{-\beta/\nu z}\,\tilde{m}(t^{1/\nu z}\epsilon).
\end{equation}
Therefore, just at criticality ($\epsilon \equiv 0$), one should observe a power-law decrease 
of the initial magnetization according to\cite{binder}
\begin{equation}\label{magescala1}
 m(t)\sim t^{-\beta/\nu z},\;\;\;\;\epsilon=0,
\end{equation}
while the logarithmic derivative of equation (\ref{magescala}) evaluated just at criticality gives\cite{albano}
\begin{equation}\label{derivescala}
\frac{\partial \ln(m(t))}{\partial\epsilon}\bigg{|}_{\epsilon\equiv 0}\sim t^{1/\nu z}.
\end{equation}
Finally, note that the scaling behavior of the cumulants
(see equations (\ref{u2}) and (\ref{u4})) can also be worked out
by using equation (\ref{escala}) and one gets
\begin{equation}
U_2(t)\sim t^{d/z},\\
\label{u2escala}
\end{equation}
\\
\noindent and
\\
\begin{equation}
U_4(t)\sim t^{d/z},
\label{u4escala}
\end{equation}
respectively.

Summing up, by properly measuring the dynamic behavior of the system at criticality one
should be able to determine the exponents, $\theta$ (equation (\ref{theta})) 
and $(d-2\beta/\nu)z$ (equation (\ref{chiescala})), by starting from disordered initial configurations; and
$\beta/\nu z$ (equation (\ref{magescala})), $1/\nu z$ (equation (\ref{derivescala})), 
and $d/z$ (equation (\ref{u2escala})), by starting from ordered initial configurations.

Usually the dimensionality of the system is known so that this set of five independent determinations 
becomes redundant in order to evaluate four critical exponents. However,
as will be shown later on, in the case of a system in a thermal gradient the complete set is 
essential for the determination of the effective dimensionality involved in
the scaling relationships. Of course, in our gradient simulations we obtained (simultaneously and by 
means of single sets of simulations)
information over a  wide range of $T$ ($T_1<T_c<T_2$), while the above discussed scaling relationships 
are expected to hold only close to
the critical region, i.e., for columns of coordinates such that
$x_c\simeq\frac{(T_c-T_1)}{(T_2-T_1)}\,L$ (see equation (\ref{gradi})).

On the other hand, equation (\ref{escala}) is also useful in order to describe the stationary scaling regime. 
In fact, for $t\rightarrow\infty$ the value of $m_0$ becomes
irrelevant and one gets the standard scaling relationship for the magnetization given by\cite{binder}
\begin{equation}\label{magestacpre}
m(\epsilon,L)=L^{-\beta/\nu}\tilde{m}(\epsilon L^{1/\nu}),
\end{equation}
where $\tilde{m}$ is a suitable scaling function. Furthermore, just at criticality ($\epsilon\equiv 0$) equation
(\ref{magestacpre}) yields
\begin{equation}\label{magestac}
m(\epsilon=0)\sim L^{-\beta/\nu},
\end{equation}
which reflects the fact that the magnetization vanishes at criticality in the thermodynamic limit ($L\rightarrow\infty$).

Also in the stationary limit, the spin-spin correlation function (equation (\ref{ge})) has its 
own scaling relationship, given by\cite{kim}
\begin{equation}\label{geescala}
g(\epsilon=0,r)\sim r^{-(d-2+\eta)},
\end{equation}
where $\eta$ is the critical exponent associated with spatial correlations, which is related to other exponents by the
relationship $\gamma=\nu(2-\eta)$\cite{kim}.

Furthermore, for homogeneous systems away from the critical
temperature, the spin-spin correlation function exhibits an exponential decay behavior given by:
\begin{equation}\label{geexpo}
g(T_i,r)\sim e^{-r/\xi_i},
\end{equation}
where $\xi_i$ is the correlation length, at temperature $T_i$, which diverges close to criticality, according to
\begin{equation}\label{xiescala}
\xi_i\sim A\,|T_i-T_c|^{-\nu},
\end{equation}
where $\nu$ is the correlation length exponent.

\section{RESULTS AND DISCUSSION}
\label{cuatro}
Let us begin our presentation with results corresponding to dynamic measurements obtained by starting 
the simulations with fully ordered
configurations such that $m(i,t=0)\equiv1,\;\forall i$, which corresponds to the ground state configuration at
$T=0$. By following the standard procedure the sample is suddenly quenched to the temperature of the heat bath
with the proper thermal gradient. So, each column of the sample evolves, at its own temperature, towards its
stationary configuration ($t\rightarrow\infty$), and in particular we focus our attention on the critical
region. According to equation (\ref{escala}), or equivalently to equation (\ref{magescala}), one expects
to measure a power-law behavior of the physical observables just at criticality, while deviations from that
type of behavior have to be found away from $T_c$. By plotting all the relevant measured 
observables ($m,\;\frac{\partial\ln(m)}{\partial(\epsilon)},\;U_2$ and $U_4$)
as obtained for samples of different sizes ($100\leq L\leq 1000$), the best set of power laws 
is obtained for $T_c=2.2691(3)$\cite{foot},
which is in full agreement with the Onsager critical
temperature\cite{onsager} of the Ising model in $d=2$. Figure 2 shows the time dependence 
of the second-order cumulant (see equations (\ref{u2}) and (\ref{u2escala})) as obtained
close to criticality by using a square lattice of side $L=1000$. Here, the thermal gradient 
is set between $T_1=2.0$ and $T_2=2.4$, while the results are averaged over $N_C=10000$
different runs. From the best fit of the data shown in figure 2 we obtained $d/z=0.474(6)$\cite{foot}.
%%%%%%%%%%%%%%%%%%%%%%%%%%%%
\vskip 1.0 true cm
\begin{figure}\label{fig2}
\centerline{
\resizebox{0.75\columnwidth}{!}{%
  \includegraphics{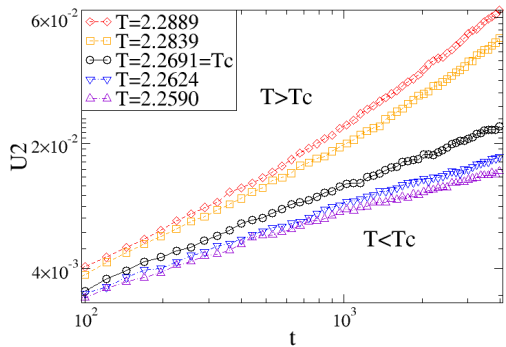}
}}
\caption{(Color online) Log-log plot of the second-order cumulant versus time as obtained for a 
lattice of side $L=1000$ and averaging over $N_C=10000$ different runs. The full line corresponds
to the best power-law behavior that is identified as the critical point. Notice that dashed (dashed-dotted) 
lines show upward (downward) deviations that are typical for $T>T_c$ ($T<T_c$).
The best fit of the solid line yields $d/z=0.474(6)$. More details in the text.}
\end{figure}
%%%%%%%%%%%%%%%%%%%%%%%%%%%

Our measurements obtained by using initially ordered configurations are completed by log-log plots of 
the time dependence of the magnetization and its logarithmic derivative, as shown
in figures 3(a) and 3(b), respectively. The best plots of the lines, obtained for data corresponding 
to the same column that was identified with the critical temperature
in figure 2, yield $\beta/\nu z=0.058(1)$\cite{foot}, and $1/\nu z=0.4857(4)$\cite{foot}, respectively.
%%%%%%%%%%%%%%%%%%%%%%%%%%%%
\vskip 1.0 true cm
\begin{figure}\label{fig3}
\centerline{
\resizebox{0.75\columnwidth}{!}{%
  \includegraphics{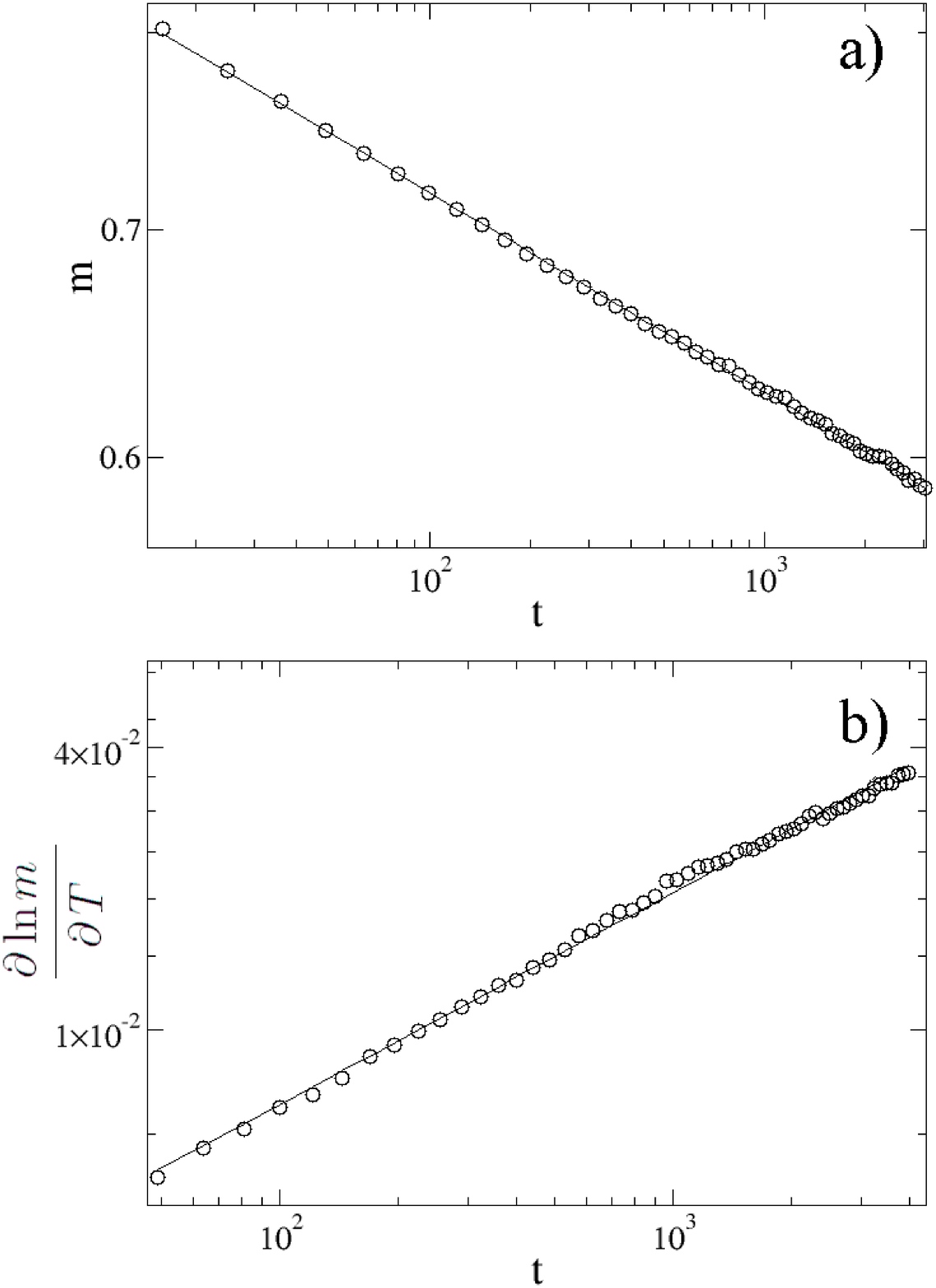}
}}
\caption{(a) (Upper panel) and (b) (lower panel) show log-log plots of the magnetization and its 
logarithmic derivative versus time, respectively. Results obtained at criticality by using samples
of side $L=1000$ and by averaging over $N_C=10000$ different runs. In both panels the straight lines
correspond to the best fits of the data shown by means of circles. The linear gradient of the thermal 
bath is set between $T_1=2.0$ and $T_2=2.4$. More details in the text.}
\end{figure}
%%%%%%%%%%%%%%%%%%%%%%%%%%%%

On the other hand, by starting the simulations with fully disordered initial configurations such that $m_0=0$,
which corresponds to samples in contact with a thermal bath at
$T=\infty$ that are then suddenly annealed to a desired thermal gradient, we measured the time evolution
of the fluctuations of the order parameter that under equilibrium
conditions is identified with the magnetic susceptibility, as shown in figure 4.
%%%%%%%%%%%%%%%%%%%%%%%%%%%%
\vskip 1.0 true cm
\begin{figure}\label{fig4}
\centerline{
\resizebox{0.75\columnwidth}{!}{%
  \includegraphics{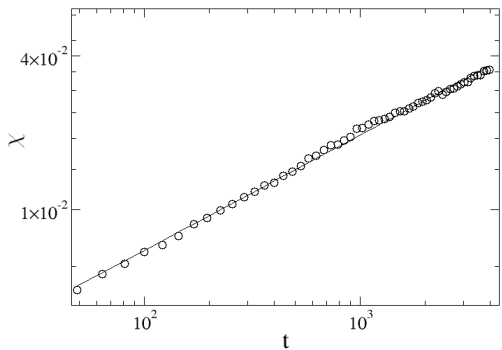}
}}
\caption{Log-log plot of the fluctuations of the order parameter versus $t$, as obtained for the 
column identified with the critical temperature in the plots shown in previous figures.
Data corresponding to $L=1000$ and taking a linear gradient set between $T_1=2.0$ and $T_2=2.4$. 
Averages are taken over $N_C=10000$ different initial configurations. The best fit of the
data, shown as a full line, has slope $\gamma/\nu z=0.3570(4)$.}
\end{figure}
%%%%%%%%%%%%%%%%%%%%%%%%%%%%
Furthermore, initially disordered configurations with a preset (vanishing) initial magnetization are suitable for the
measurement of the initial increase of the order parameter, which after the
scaling arguments discussed within the context of equation (\ref{theta}), is expected to be observed for times such
 that $t<<m_0^{-z/x_0}$. Figure 5 shows log-log plots of $m$
versus $t$ as obtained for the different values of $m_0$ ($0.035\leq m_0\leq 0.050$). The data shown in figure 5
 correspond to the previously identified "critical" column.
The best fits of the curves give estimates of the initial increase exponent $\theta$ that,
after a proper extrapolation to the limit $m_0\rightarrow 0$ (not shown here for the sake of space),
yields $\theta=0.196(6)$\cite{foot}.
%%%%%%%%%%%%%%%%%%%%%%%%%%%%
\vskip 1.0 true cm
\begin{figure}\label{fig5}
\centerline{
\resizebox{0.75\columnwidth}{!}{%
  \includegraphics{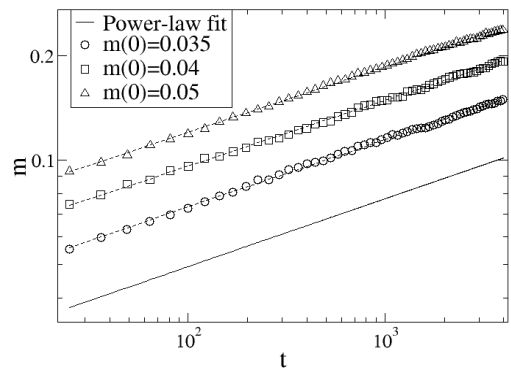}
}}
\caption{Log-log plot of the magnetization versus $t$, as obtained for the column identified with the critical
temperature in the plots shown in previous figures.
Data corresponding to $L=1000$ and taking a linear gradient set between $T_1=2.0$ and $T_2=2.4$. Averages are
taken over $N_C=10000$ different initial configurations.
Each curve corresponds to a different initial magnetization $m_0$, as indicated, and the dashed lines are the fits
of the data. The full straight line corresponds to $\theta=0.196$, i.e. the value of the initial increase exponent
obtained by extrapolation to $m_0\rightarrow 0$. More details in the text.}
\end{figure}
%%%%%%%%%%%%%%%%%%%%%%%%%%%%
Based on the results obtained by means of dynamic measurements only, we are in a condition to outline a few
interesting preliminary conclusions on the critical behavior of the Ising
magnet in contact with a gradient thermal bath. On the one hand, the best
fits of all physical observables were found at the same column (i.e., the same temperature) of the sample, which
has been identified as the critical temperature $T_c=2.2691(3)$\cite{foot}. This value is in full agreement
with the well known exact result early
evaluated by Onsager, i.e., $T_{Onsager}\simeq 2.2692$.
%when statistical errors are properly considered.
On the other hand, by using $\beta/\nu z=0.058(1)$ and $1/\nu z=0.4857(4)$\cite{foot} as determined by means of the
measurements of the magnetization and its logarithmic derivative, respectively,
we obtained $\beta=0.120(2)$\cite{foot}, in excellent agreement with the exact value $\beta=1/8=0.125$\cite{onsager}.
Additionally, by assuming that the hyperscaling relationship $d\nu-2\beta=\gamma$ holds,
just dividing by $\nu z$ one can write
\begin{equation}\label{hiper}
\frac{d}{z}-\frac{2\beta}{\nu z}-\frac{\gamma}{\nu z}=0,
\end{equation}
where equation (\ref{hiper}) can be interpreted as a "dynamical" hyperscaling relationship. Then, by replacing
the measured exponents in
equation (\ref{hiper}) we obtain $0.001(9)$ on the
right-hand of the equality, a result that strongly supports the validity of hyperscaling. On the other hand, we
found that the exponent for the initial increase of the order parameter ($\theta=0.196(6)$)
is in full agreement
with previous numerical results obtained by applying the Metropolis dynamics to the two-dimensional Ising magnet
($d=2$) in a homogeneous bath, namely, $\theta=0.191(1)$\cite{okano}. However, it should be mentioned that this
exponent depends on the dynamics used (e.g. Metropolis, Glauber, Heat-Bath, etc.)
and that our Gradient-Metropolis dynamics may not
give the same exponents as the proper standard Metropolis dynamics.

Let us now point our attention to stationary results obtained after disregarding $N_D=5\times10^5$ Monte Carlo
steps, in order to allow for the stabilization of the sample, and evaluated during the
subsequent time interval of $5\times10^5$ Monte Carlo steps.

Figure 6(a) shows plots of magnetization profiles (i.e., plots of $m(i)$ versus $T_i$, where $T_i$ is the temperature
of the $i$th column), as obtained for samples of different
sizes. Since the temperatures at the extremes of the sample are kept constant at $T_1=0.64$ and $T_2=3.40$, 
each curve in figure 6(a) corresponds to different gradients. It follows
that finite-size effects are negligible in the low-temperature range, e.g., for $T<2.1$,
while these effects become slightly evident above criticality.
%%%%%%%%%%%%%%%%%%%%%%%%%%%%
\vskip 1.0 true cm
\begin{figure}\label{fig6}
\centerline{
\resizebox{0.75\columnwidth}{!}{%
  \includegraphics{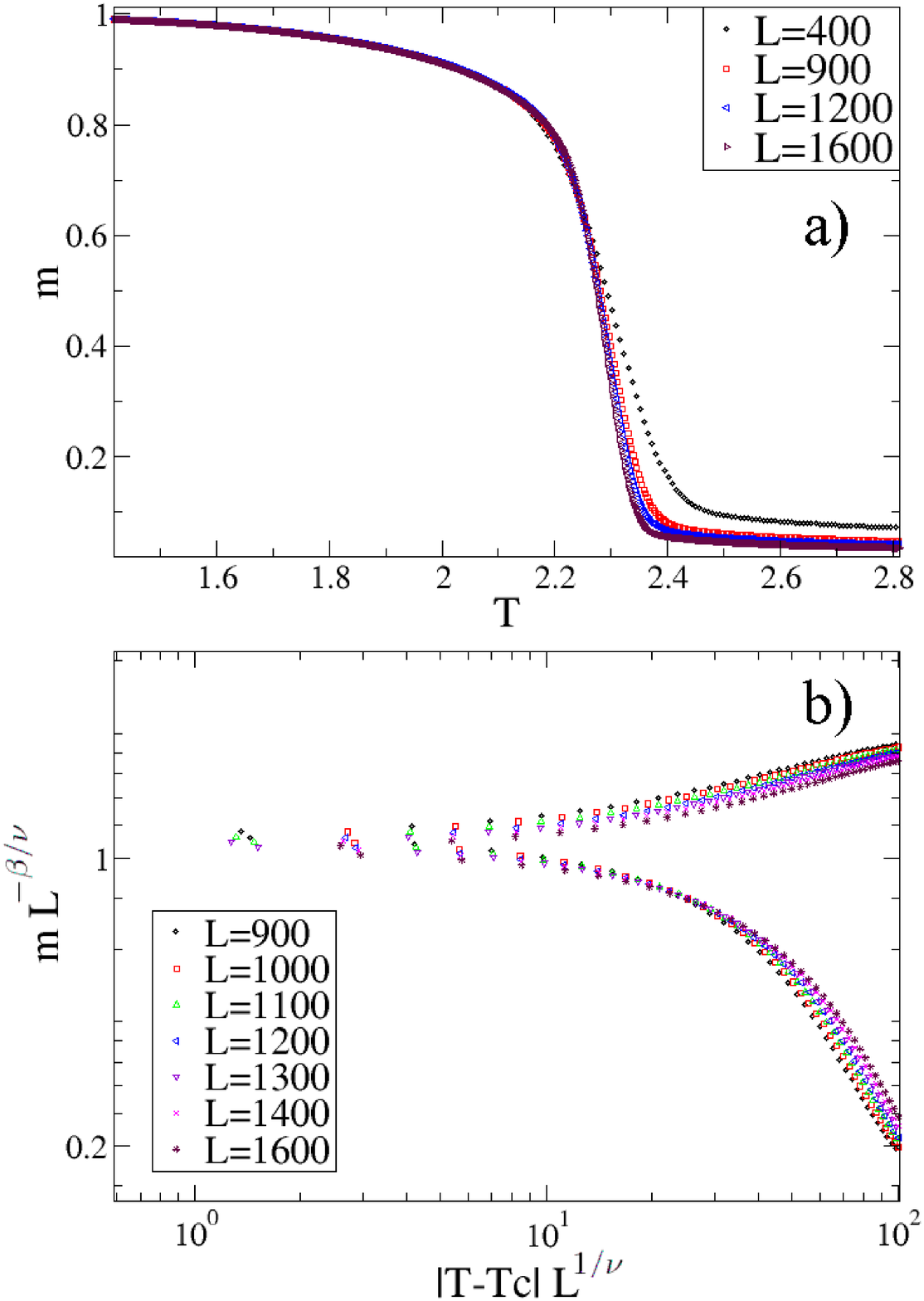}
}}
\caption{(Color online) (a) Plot of the magnetization profiles versus $T$ as obtained for samples 
of different thermal gradients. The temperatures at the extremes of the sample are $T_1=0.64$ and
$T_2=3.40$, and results are averaged over $N_D=5\times10^5$ Monte Carlo steps after disregarding 
the initial $5\times10^5$ Monte Carlo steps in order to allow for the stabilization of the
samples. (b) Scaling plots
of the data already shown in (a) as obtained by using equation (\ref{magestacpre}). More details in the text.}
\end{figure}
%%%%%%%%%%%%%%%%%%%%%%%%%%%%
It is worth mentioning that the results shown in figure 6(a) can be replotted in order to test whether the
standard scaling also holds for gradient measurements.
 In fact, figure 6(b) shows log-log plots of $m(T_i,L)L^{\beta/\nu}$ versus
 $|T_i-T_c|L^{1/\nu}$ as obtained by using the data already shown in figure 6(a). The
quality of the collapse supports the validity of the standard scaling Ansatz for the case of our gradient measurements;
 however, the terms of correction to scaling, which are neglected in our analysis, cannot be disregarded.
 On the other hand, just by selecting the
critical column one has that equation (\ref{magestac}) simply gives the monotonic decay of the magnetization 
when the system size is increased, namely $m(T=T_c)\sim L^{-\beta/\nu}$,
a result that has been verified in figure 7 where the best fit of the data 
corresponds to $\beta/\nu=0.125(9)$\cite{foot}.
%%%%%%%%%%%%%%%%%%%%%%%%%%%%
\vskip 1.0 true cm
\begin{figure}\label{fig7}
\centerline{
\resizebox{0.75\columnwidth}{!}{%
  \includegraphics{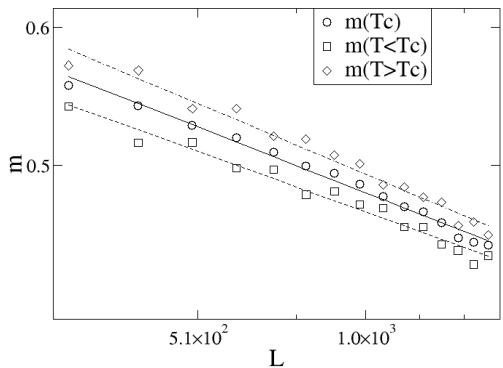}
}}
\caption{Log-log plot of the magnetization, measured close to criticality, versus the system side $L$. 
Data corresponding to the critical column (open circles), and two adjacent
columns: $T<T_c$, open squares, and $T>T_c$, open diamonds, respectively. Results obtained by 
averaging over $5\times10^5$ Monte Carlo steps after disregarding the first $N_D=5\times10^5$ Monte
Carlo steps in order to allow for the stabilization of the system. The full line has slope $-\beta/\nu=0.125(9)$, 
and corresponds to the best fit of the data. The other two lines
correspond to the fits of the data from adjacent columns and give the error bars to the result\cite{foot}. More details
in the text.}
\end{figure}
%%%%%%%%%%%%%%%%%%%%%%%%%%%%
As already established in the field of critical phenomena, the so-called cumulants
(see equations (\ref{u2}) and (\ref{u4})) are suitable functions of the moments of the order
parameter distribution function whose pre-scaling factor, disregarding high-order finite-size scaling corrections,
is independent of the system size. So,
plots of the cumulants versus the control parameter (i.e., the temperature of the gradient thermal bath)
 should exhibit a common intersection point,
 as it is indicated with full straight vertical lines in the data shown
in figures 8(a) and  8(b), obtained with our gradient system for $U_2(L,T)$ and $U_4(L,T)$,
respectively. A careful inspection of the critical region close to the
interaction points reveals very small but systematic shifts of the intersection points between
 curves corresponding to adjacent system sizes. After proper extrapolation of the intersection points to the
thermodynamic limit (not shown here for the sake of space), we obtain $T_c(\infty)=2.284(1)$\cite{foot}
 and $T_c(\infty)=2.283(3)$\cite{foot}, for
 data corresponding to $U_2$ and $U_4$, respectively.
%%%%%%%%%%%%%%%%%%%%%%%%%%%%
\vskip 1.0 true cm
\begin{figure}\label{fig8}
\centerline{
\resizebox{0.75\columnwidth}{!}{%
  \includegraphics{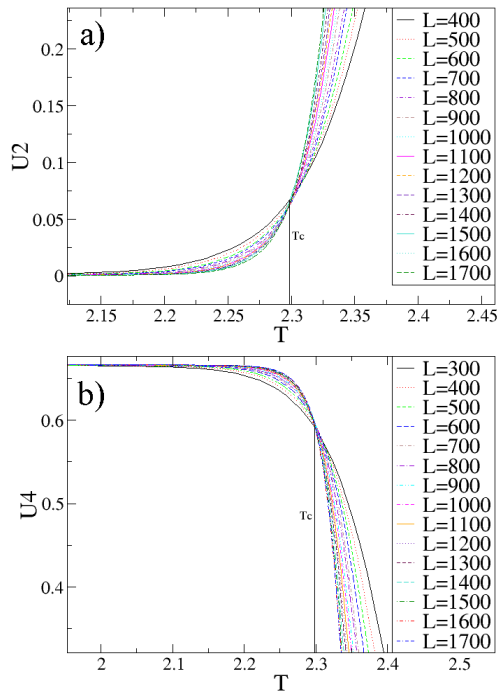}
}}
\caption{(Color online) Plots of the second- (a) and fourth-order (b) cumulants $U_2$ and $U_4$ versus the
 temperature of the column, as obtained for different values of the system side $L$ as indicated.
The vertical straight lines
on each graph indicate the intersection points and are shown for the sake of comparison.}
\end{figure}
%%%%%%%%%%%%%%%%%%%%%%%%%%%%

On the other hand, plots of the fluctuations of the order parameter, which are identified with 
the susceptibility in standard measurements, show peaks close to criticality as expected
(see figure 9). In fact, it is known that the susceptibility exhibits rounding and shifting 
effects as the size-dependent "critical" temperature of the peaks
($T_c(L)$) converges toward the true critical temperature according to a standard finite-size 
scaling relationship given by\cite{binder}
\begin{equation}\label{tescala}
T_c(L)=T_c(\infty)+constant\,L^{-1/\nu}.
\end{equation}
So, the inset of figure 9 shows plots of $T_c(L)$ versus $L^{-1}$ (here we assume $\nu=1$ for the 
correlation length exponent as justified below) that
yield $T_c(\infty)=2.296(1)$\cite{foot}.
%%%%%%%%%%%%%%%%%%%%%%%%%%%%
\vskip 1.0 true cm
\begin{figure}\label{fig9}
\centerline{
\resizebox{0.75\columnwidth}{!}{%
  \includegraphics{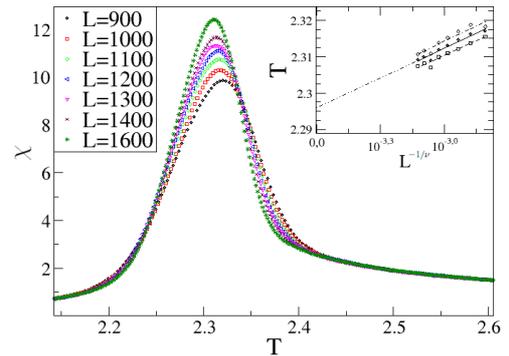}
}}
\caption{(Color online) Plots of the fluctuations of the magnetization versus the temperature of the 
column, obtained by means of simulations with systems of different side,
as listed in the figure.
The inset shows the extrapolation of the effective "critical" temperature according to 
equation (\ref{tescala}), which yields $T_c(\infty)=2.296(1)$ as the best fit (open circles).
Notice that extrapolation of data corresponding to the columns adjacent to the critical one are also 
included, such that open diamonds and open squares correspond to $T<T_c$ and $T>T_c$, respectively. More details in
the text.}
\end{figure}
%%%%%%%%%%%%%%%%%%%%%%%%%%%%

In additional simulations, we studied the spin-spin correlation functions, both parallel and
perpendicular to the gradient direction.
We measured the correlation function in the perpendicular direction, by using equation (\ref{ge})
for different temperatures, both lower and higher than $T_c\simeq 2.26918$. We used a 
rectangular lattice of size $L=300$, $M=4000$, and
set the temperatures of the extremes of the sample at $T_1=0.8$ and $T_2=3.74$, respectively. 
For each measurement  corresponding to a certain
temperature outside criticality, we plotted the correlation
function versus the distance $r$, in log-linear plots as shown in figure 10(a). Then, by fitting
the data to an exponential decay, according to equation (\ref{geexpo}),
we obtained the values of the correlation length $\xi_i(|T_i-T_c|)$.
The inset of figure 10(a) shows a linear-linear plot of $\xi(T)$ versus $1/(T-T_c)$ that yields a straight line.
This behavior confirms that equation (\ref{xiescala}) holds with $\nu=1$, and from the slope we 
also obtained the prefactor given by $A=1.16(4)$.
 %%%%%%%%%%%%%%%%%%%%%%%%%%%%
\vskip 1.0 true cm
\begin{figure}\label{fig10}
\centerline{
\resizebox{0.75\columnwidth}{!}{%
  \includegraphics{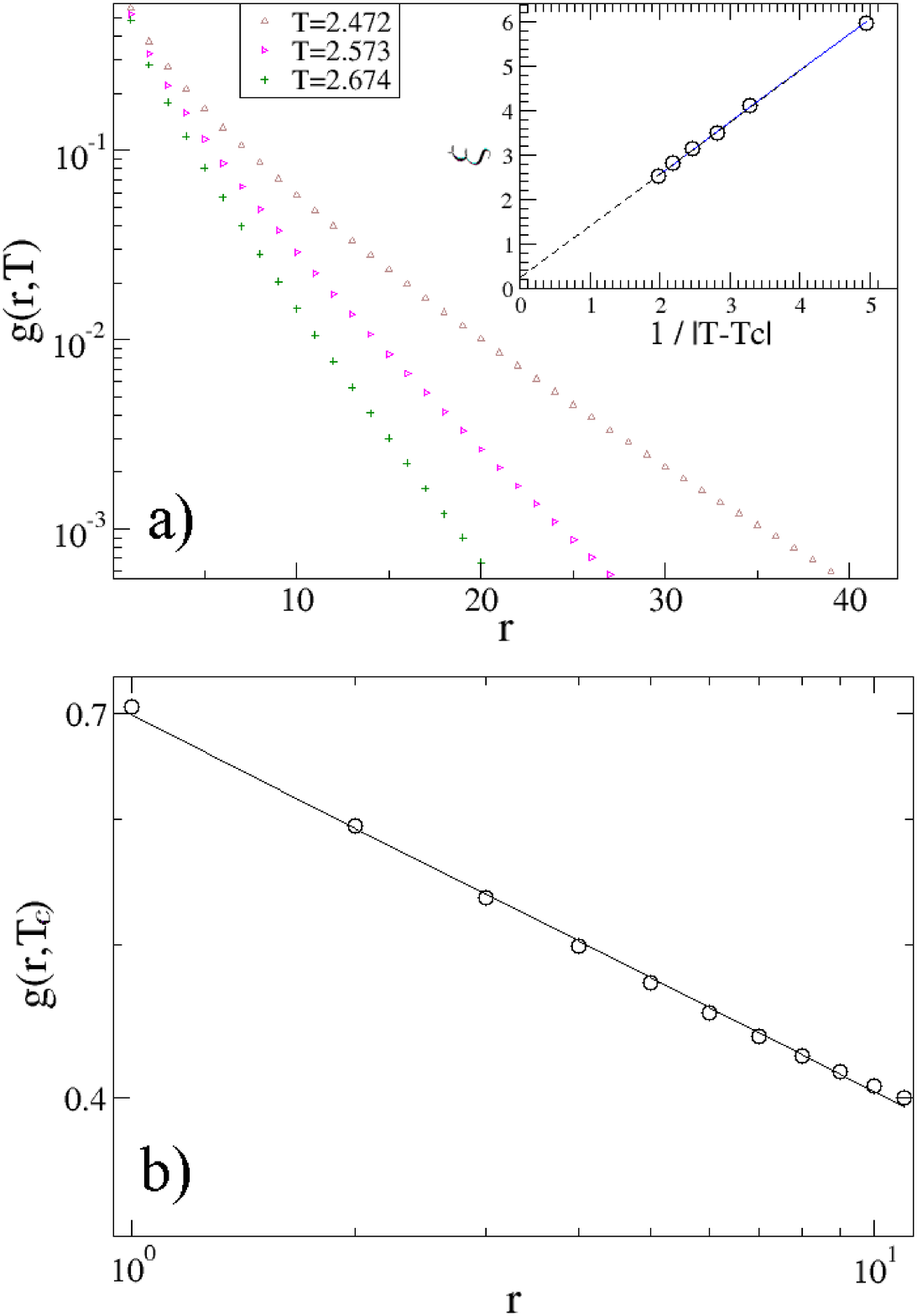}
}}
\caption{(Color online) (a) Log-linear plots of the vertical spin-spin correlation function $g(T,r)$
versus the distance $r$. Different
curves correspond to different temperatures away from criticality, as indicated. For each curve an exponential decay fit
was made, obtaining values
of the correlation length $\xi(T)$. The inset in (a) shows the values of $\xi$ versus $1/|T-T_c|$.
The linear fit indicates that equation (\ref{xiescala}) holds, with an
exponent $\nu=1$. Furthermore, the slope of the fit gives us the
value of the pre factor $A$ in equation (\ref{xiescala}), which is found to be $A=1.16(4)$.
(b) Log-log plot of $g(T,r)$ versus $r$, as obtained for the
 critical temperature, with its corresponding power-law fit (full line). The exponent 
obtained corresponds to $d-2+\eta=0.238(3)$.
 A rectangular lattice was used, with $L=300$ and $M=4000$.
 The temperature gradient was set between $T_1=0.8$ and $T_2=3.74$. Results obtained by averaging
 over $5\times10^5$ Monte Carlo steps after disregarding the first $N_D=5\times10^5$ Monte
Carlo steps in order to allow for the stabilization of the system. More details in the text.}
\end{figure}
%%%%%%%%%%%%%%%%%%%%%%%%%%%%
On the other hand, for the correlation function corresponding to the column of the system with temperature closest
 to $T_c\simeq 2.26918$, we performed a power-law fit versus the distance $r$, according to equation (\ref{geescala})
 (see figure 10(b)),
which yields
$d-2+\eta=0.238(3)$. By eliminating the value of $d$ from the exponent equation with the use of relationship
$d\nu-2\beta=\gamma$, and by using values already calculated from both stationary and dynamic simulations,
we obtained $\eta=1.22(1)$.
This value is in excellent agreement with $\eta=5/4=1.25$, found by taking
$d=1$ in the relationship that results from combining $\gamma=\nu(2-\eta)$
and $d\nu-2\beta=\gamma$ with the
exact exponent values of the two-dimensional Ising Model, i.e., $\nu=1$ and $\beta=1/8$\cite{onsager}.

On the other hand, the correlation function in the direction parallel to the thermal gradient
has also been measured, by using a system of $L=5000$ and $M=100$, with a gradient between
 temperatures $T_1=0.8$ and $T_2=3.74$, respectively.
The correlation function studied corresponds to $g(T,r)$ at the critical temperature, namely, $g(T_c,r)$.
It is worth mentioning that correlations along the direction of the applied gradient involve spins
at different temperatures, so a quantitative study such as the one done for vertical correlations is no longer possible.
Instead, we gain an insight into the qualitative behavior of the Ising ferromagnet in a thermal gradient. In fact,
figure 11 shows
$g(T_c,r)$ versus the distance $r$. For $r>0$, we found that correlations between the spins in the critical column
and those placed in columns at higher temperatures are almost negligible. On the other hand, 
for $r<0$ correlations between the spins located in the critical column
and those placed in columns at lower temperatures are measured. Here, the correlations first
abruptly decrease and then they become almost constant for colder temperatures. The inset in figure 11 shows a log-log
 plot of a zoomed view of the area near criticality, where absolute values of $r$ were taken, in order
to account for both branches of $g(r)$, namely, for $r>0$ and $r<0$, respectively.
 The full line is a plot of a power-law decay with exponent
 $1/4=0.25$ and is shown for the sake of comparison. By comparing the measured points with
 the full line, it can be inferred that
 close enough to the critical column correlations have a functional behavior in agreement with the well
 known scaling law $r^{-(d-2+\eta)}$. This result reflects the fact that for the considered gradient
  $\frac{\Delta T}{\Delta L}=0.588\times 10^{-3}$, the region close to $r=0$ is almost at the critical temperature.
%%%%%%%%%%%%%%%%%%%%%%%%%%%%
\vskip 1.0 true cm
\begin{figure}\label{fig11}
\centerline{
\resizebox{0.75\columnwidth}{!}{%
  \includegraphics{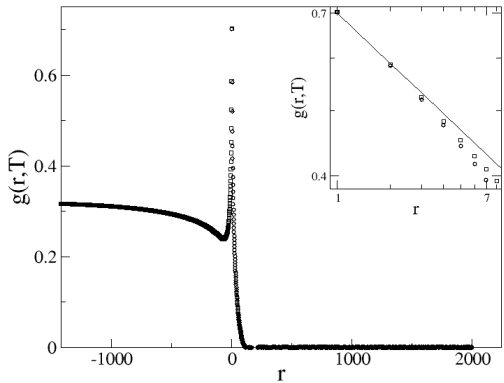}
}}
\caption{Plots of the correlation function $g(T,r)$, where the origin is taken just at the critical temperature.
Here, for $r<0$ ($r>0$) one has
$T<T_c$ ($T>T_c$).
 A rectangular lattice was used, with $L=5000$ and $M=100$.
 The temperature gradient was set between $T_1=0.8$ and $T_2=3.74$, respectively. Results obtained by
averaging over $5\times10^5$ Monte Carlo
 steps after disregarding the first $N_D=5\times10^5$ Monte
Carlo steps in order to allow for the stabilization of the system. The inset shows an enlarged view of the area
close to criticality, on log-log scale. Squares (circles) correspond to correlations with spins in the direction of
$T<T_c$ ($T>T_c$). The full line is a plot of the power law of $r$ with
 exponent $1/4=0.25$ and is shown for the sake
 of comparison. Absolute values of $r$ have been taken, in order to show both branches of $g(r)$.
More details in the text.}
\end{figure}
%%%%%%%%%%%%%%%%%%%%%%%%%%%%

Let us now discuss the results obtained in the present paper and perform comparisons to 
existing numerical results and/or exact values corresponding to the Ising magnet
in a homogeneous thermal bath. Table I summarizes over seven independent estimations of the critical 
temperature as follows from the measurement of different
physical observables in our gradient Ising system. In all cases we obtained an excellent agreement 
with the exact value, pointing out that the proposed gradient method is suitable
for accurate determinations of critical points.

\begin{center}
\begin{table}
\begin{tabular}{ | l | l | l | }
\hline
Type & Observable & $T_c$\\ \hline
Dynamic & Magnetization & 2.2691(3)\\ \hline
Dynamic & Cumulant $U_2$ & 2.2691(3)\\ \hline
Dynamic & $2^{nd}$ order magnetization & 2.2691(3)\\ \hline
Stationary & Cumulant $U_2$ &  2.284(1)\\ \hline
Stationary & Cumulant $U_4$ & 2.283(3)\\ \hline
Stationary & Scaling of $m$ & 2.282(3)\\ \hline
%Stationary & Extrapolation of $m(T_c)$ & \\ \hline
Stationary & Extrapolation of $T_c(L)$ from $\chi$ & 2.296(1)\\ \hline
Exact\cite{onsager} & $\;\;\;\;\;\;\;\;\;\;$- & $\sim 2.26918$ \\ \hline
\end{tabular}
\caption{List of critical temperatures (third column) obtained by means of dynamic and stationary measurements, 
as listed in the first column. More details in the text.}
\end{table}
\end{center}

Pointing now our attention to the measured critical exponents, we have already shown that our 
dynamic measurements of $d/z$, $\beta/\nu z$, and $\gamma/\nu z$ are fully consistent
with a sort of "dynamic" hyperscaling relationship (equation (\ref{hiper})). However, since 
the dimensionality entering in equation (\ref{hiper}) appears in a quotient
between exponents, more precisely as $d/z$ as follows from dynamic measurements of the cumulant, 
one needs independent measurements in order to
obtain an unbiased estimation of $d$. The same reasoning applies to the dynamic exponent $z$.
Therefore, we conclude that in order to obtain all the critical exponents we have to combine measurements 
corresponding to both the short-time dynamic
simulations, namely, $d/z$, $\beta/\nu z$, and $\gamma/\nu z$, and stationary simulations,
 namely, $\beta/\nu$ and $d-2+\eta$. The results are shown in Table II, along with the values
for the exponents that are known exactly ($\beta$, $\nu$, $\eta$, and $\gamma$), and the best 
available numerical results ($z$ and $\theta$), for the two-dimensional
Ising model with a homogeneous thermal bath. It is worth mentioning that the error bars in each value
were propagated from those of the original exponents, which in turn were calculated considering the
values of the physical observables from the adjacent columns (at different temperatures) 
to the critical one\cite{foot}.

\begin{center}
\begin{table}
    \begin{tabular}{ | l | l | l | }
    \hline
    Exponent & Present work & Exact/numerical value\\ \hline
    $\beta\;$ & 0.120(2) & 1/8=0.125 \\ \hline
    $\nu\;$ & 0.96(7) & 1 \\ \hline
    $\eta\;$ & 1.238(3) & 5/4=1.25 (Details in the text)\\ \hline
    $\gamma\;$ & 0.736(1) & 3/4=0.75 (Details in the text)\\ \hline
    $z\;$ & 2.16(4) & 2.166(7)\cite{wang}\\ \hline
    $\theta\;$ & 0.196(6) & 0.191(1)\cite{okano}\\ \hline
    $d\;$ & 1.02(8) & $-$ \\ \hline
    \end{tabular}
\caption{List of the critical exponents (second column), along with the value of the dimensionality, 
obtained by combining dynamic and stationary measurements.
The third column lists the exact value, in the first four rows, and the comparison value 
from numerical simulations, in the fifth and sixth rows.}
\end{table}
\end{center}

There is an interesting result that involves both $\gamma$ and $\eta$ because the values obtained are different
in a quantity of $1$ as compared with the exact values for the Ising ferromagnet for $d=2$, 
namely $\gamma=7/4$ and $\eta=1/4$.
Furthermore, if we determine the values of $\gamma$ and $\eta$ from the scaling relationships $d\nu-2\beta=\gamma$
and $\gamma=\nu(2-\eta)$ by using the values of the exponents found in this work, we roughly reach the same
results, namely $\gamma=0.74(8)$ and $\eta=1.23(8)$ respectively.
The reason for this apparent disagreement becomes evident when we calculate, from the combination of
dynamic and stationary results, the value
of the effective dimensionality, obtaining $d=1.02(8)$, which strongly suggests $d=1$ as judged by the 
uncertainty interval. If we now
use $d=1$ and re-calculate the exponents of the Ising ferromagnet, we obtain $\gamma=3/4=0.75$ and
$\eta=5/4=1.25$, which are in perfect agreement with the exponents found by means of our simulations. This
value for the effective dimensionality $d$ takes into account the fact that due to the presence of a thermal gradient,
the thermodynamic functions are calculated as profiles
of columns, thus lowering the effective dimensionality exactly by one unit.

\section{CONCLUSIONS}
\label{cinco}
We have studied, by means of Monte Carlo numerical simulations, the dynamic and stationary critical behavior 
of the two-dimensional Ising ferromagnet, in contact with a thermal bath that exhibits a linear gradient between 
low- and high- temperature extremes at  $T_1$ and $T_2$ ($T_2>T_1$), respectively. It is
worth mentioning that our proposed approach, is based on the assumption that the ferromagnet is in thermal equilibrium
with the heat bath that intrinsically has a thermal gradient. This is in contrast to previous studies by other 
authors \cite{dgrad0,dgrad1,dgrad2,dgrad3} who placed only the extremes of the magnet
in contact with two thermal baths at $T_1$ and $T_2$, and subsequently measured the thermal conduction between 
these baths. While for the numerical implementation of this "conductivity approach" one needs to use a 
suitable "Creutz demon" algorithm\cite{creutz}, our more straightforward approach can be implemented by means
of standard algorithms, e.g., the Metropolis dynamics, just as in the present paper. Of course, our whole sample 
is out of equilibrium, but a key feature is that each column of the magnet (or sample in general) can be 
considered in equilibrium with the gradient thermal bath that is in contact with it.
In this way, now one not only has information of the system under study for a wide range of temperatures
($T_1<T<T_2$), but also by means of a suitable choice of the temperatures of the extremes of the sample, 
such that $T_1<T_c<T_2$, where $T_c$ is the critical temperature, one can deal with samples simultaneously 
exhibiting the ordered and the disordered phases. In this way, instead of measuring average values of
the physical observables at each $T$, one actually measures "thermal profiles" of the observables averaged 
over ($d-1$)-dimensional columns of the sample.
For this reason, the effective dimensionality entering in the scaling relationships is $d=1$.

Focusing our attention on the dynamic measurements, the measured thermal profiles allow us to 
simultaneously follow the time evolution of the sample for
all temperatures of interest and, by drawing suitable plots, quickly identify the critical temperature 
(corresponding to a particular column) and determine the relevant critical
 exponents. Similarly, for the case of stationary measurements, we take advantage of the wide information 
available (critical temperature, critical exponents, correlation functions, etc.) that can be obtained 
just by performing a single simulation. The results obtained for the critical temperature and critical exponents
are summarized and discussed in the context of Tables I and II, respectively. Based on that analysis, 
we conclude that the figures are remarkably accurate when one considers the computational effort involved. 
In fact, full agreement is obtained when our results are compared with exactly known values, e.g.,
for the critical temperature $T_c$, and the critical exponents $\beta$, $\nu$, $\gamma$, and $\eta$, 
as well as with the best available data taken from numerical simulations of other authors, e.g., 
for the exponents $\theta$ and $z$.

Based on the results presented and discussed in this paper, we conclude that the study of critical systems
in the presence of a gradient of the suitable control parameter
is a useful and powerful tool to gain rapid inshight on the behavior of the system, which
could eventually be studied in further details by using more sophisticated methods . 
Furthermore, the studies of material systems under gradient conditions could shed light on the understanding 
of a wide variety of experimental and technologically useful situations.

{\bf  ACKNOWLEDGMENTS}. This work was financially supported by CICPBA, CONICET, UNLP and ANPCyT (Argentina).

%and \cite{RefJ}
%\subsection{Subsection title}
%\label{sec:2}
%as required. Don't forget to give each section
%and subsection a unique label (see Sect.~\ref{sec:1}).
%
% For one-column wide figures use
%\begin{figure}
% Use the relevant command for your figure-insertion program
% to insert the figure file.
% For example, with the option graphics use
%\resizebox{0.75\columnwidth}{!}{%
%  \includegraphics{example.eps}
%}
% If not, use
%\vspace{5cm}       % Give the correct figure height in cm
%\caption{Please write your figure caption here}
%\label{fig:1}       % Give a unique label
%\end{figure}
%
% For two-column wide figures use
%\begin{figure*}
% Use the relevant command for your figure-insertion program
% to insert the figure file. See example above.
% If not, use
%\vspace*{5cm}       % Give the correct figure height in cm
%\caption{Please write your figure caption here}
%\label{fig:2}       % Give a unique label
%\end{figure*}
%
% For tables use
%\begin{table}
%\caption{Please write your table caption here}
%\label{tab:1}       % Give a unique label
% For LaTeX tables use
%\begin{tabular}{lll}
%\hline\noalign{\smallskip}
%first & second & third  \\
%\noalign{\smallskip}\hline\noalign{\smallskip}
%number & number & number \\
%number & number & number \\
%\noalign{\smallskip}\hline
%\end{tabular}
% Or use
%\vspace*{5cm}  % with the correct table height
%\end{table}
%
% BibTeX users please use
% \bibliographystyle{}
% \bibliography{}

\begin{thebibliography}{99}
%
% and use \bibitem to create references.
%
%\bibitem{RefJ}
% Format for Journal Reference
\bibitem{dgrad0} K. Saito, S. Takesue, and S. Miyashita, Phys. Rev. E. {\bf 59}, (1999) 2783. 

\bibitem{dgrad1} M. Neek-Amal, R. Moussavi, and H. R. Sepangi, Physica A  {\bf 371}, (2006) 424.

\bibitem{dgrad2} M. Casartelli, N. Macellari, and A. Vezzani, Eur. Phys. J. B {\bf 56}, (2007) 149.
 
\bibitem{dgrad3} E. Agliari, M. Casartelli, and A. Vezzani, Eur. Phys. J. B {\bf 60}, (2007) 499.

\bibitem{jsm} E. Agliari, M. Casartelli and A. Vezzani, J. Stat. Mech., P07041 (2009).

\bibitem{isingorig} E. Ising, Zeits. F. Physik {\bf 31}, (1925) 253.

\bibitem{tanaka} M. Tanaka, T. Wakamatsu, and H. Kobayashi, IEEE Trans. J. Magn. Japan {\bf 2}, (1987) 728.

\bibitem{aleman} M. M. Schwickert, K. A. Hannibal, M. F. Toney, M. Best, L. Folks, J.-U. Thiele, A. J. Kellock,
and D. Weller, J. Appl. Phys. {\bf 87}, (2000) 6956.

\bibitem{xiong} L. YoongXiong, E. Ahmad, Y. Xu and K. Barmak, IEEE Trans. Magn. {\bf 41}, (2005) 3328.

\bibitem{creutz} M. Creutz, M, Phys. Rev. Lett. {\bf 50}, (1983) 1411.

\bibitem{KS}  L. Kadanoff and J. Swift, Phys. Rev. {\bf 165}, (1968) 310.

\bibitem{Q2R} G. Y. Vichniac, Physica D, {\bf 10}, (1984) 96.

\bibitem{kit} Kittel, C., \emph{Introduction to Solid State Physics}, Eighth ed., 
John Wiley {\&} Sons, Inc; USA (2005)

\bibitem{albano} E. V. Albano, M. A. Bab, G. Baglietto, R. A. Borzi,
T. S. Grigera, E. S. Loscar, D. E. Rodriguez, M. L. Rubio Puzzo, and G. P. Saracco, 
Rep. Prog. Phys. {\bf 74}, (2011)  026501.

\bibitem{onsager} L. Onsager, Phys. Rev. {\bf 65}, (1944) 117.

\bibitem{foot0} It is worth mentioning that by setting $\epsilon\equiv 0$ the dependence of 
the observables on the horizontal coordinate along the thermal gradient is supresed.

\bibitem{janssen} H. K. Janssen, B. Schaub, and B. Schmittmann, Z. Phys. B {\bf 73} (1989) 539.

\bibitem{binder} K. Binder, Rep. Prog. Phys. {\bf 60}, (1997) 487.

\bibitem{kim} K. Christensen, and N. R. Moloney, \emph{Complexity and Criticality}, 
(Imperial College Press, London, 2005).

\bibitem{foot} It is worth mentioning that the error bars
only account for the changes in the values of the physical observables when adjacent columns 
are considered, and an error of the order of $(T_{i+1}-T_{i-1})/L$ is introduced in the temperature.
So, our reported error bars have to be enlarged when considering both statistical errors and scaling 
corrections that are ignored.

\bibitem{okano} K. Okano, L. Sch\"{u}lke, K. Yamagishi, and B. Zheng, Nucl. Phys. B {\bf 485}, (1997) 727

\bibitem{wang} F. G. Wang, and C. K. Hu, Phys. Rev. E {\bf 56}, (1997) 2310


%Author, Journal \textbf{Volume}, (year) page numbers.
% Format for books
%\bibitem{RefB}
%Author, \textit{Book title} (Publisher, place year) page numbers
% etc
\end{thebibliography}
%
% Non-BibTeX users please use

\end{document}